\documentclass[published]{JHEP3} 

\JHEP{00(2002)000}

\JHEPspecialurl{http://jhep.sissa.it/JOURNAL/JHEP3.tar.gz}

\usepackage{epsfig,multicol,bbm}

\newcommand\fverb{\setbox\pippobox=\hbox\bgroup\verb}
\newcommand\fverbdo{\egroup\medskip\noindent%
                        \fbox{\unhbox\pippobox}\ }
\newcommand\fverbit{\egroup\item[\fbox{\unhbox\pippobox}]}
\newbox\pippobox

\def\a{\alpha}
\def\b{\beta}
\def\g{\gamma}
\def\d{\delta}
\def\D{\Delta}

\def\t{\theta}

\def\l{\lambda}

\def\be{\begin{equation}}
\def\ee{\end{equation}}
\def\ba{\begin{eqnarray}}
\def\ea{\end{eqnarray}}

\newcommand{\rpar}{\stackrel{\leftarrow}{\partial}}
\newcommand{\lpar}{\stackrel{\rightarrow}{\partial}}

\newcommand{\no}{\nonumber}

\title{Nambu-like odd bracket on Grassmann algebra}

\author{Dmitrij V. Soroka and Vyacheslav A. Soroka\\
        Kharkov Institute of Physics and Technology\\
        1 Akademicheskaya St., 61108 Kharkov, Ukraine\\
        E-mail: \email{dsoroka@kipt.kharkov.ua}, \email{vsoroka@kipt.kharkov.ua}}
\received{}             
\revised{}
\accepted{}             

\preprint{\hepth{9912999}}      

\abstract{The Grassmann-odd Nambu-like bracket corresponding to an arbitrary 
Lie algebra and realized on the Grassmann algebra is proposed.}


\dedicated{Dedicated to Nikita Dmitriyevich Soroka.\\}

\begin{document} 


\section{Introduction}

In the paper \cite{ss1} the Grassmann-odd Nambu \cite{n} bracket corresponding 
to the $SO(3)$ group and built up only of the Grassmann variables $\t_\a$ 
has been introduced. The purpose of the present paper is to extend this 
construction onto an arbitrary Lie group.

In \cite{s1,s2,ss2,ss3} the linear odd bracket corresponding to an 
arbitrary Lie group and realized solely on the Grassmann variables has been
proposed
\ba\label{1.1}
\{A,B\}_1=A\rpar_{\t_\a}{c_{\a\b}}^\g\t_\g\lpar_{\t_\b}B,
\ea
where ${c_{\a\b}}^\g$ are structure constants,
$\rpar$ and $\lpar$ are the right and left derivatives and 
$\partial_{\t_\a} \equiv {\partial \over {\partial \t_\a}}$. It was constructed
for this bracket at once three Grassmann-odd nilpotent $\D$-like differential 
operators of the first, second and third orders with respect to the Grassmann 
derivatives
\ba\label{1.2}
\D_{+1}=\t^\a\t^\b{c_{\a\b}}^\g\partial_{\t^\g},
\ea
\ba\label{1.3}
\D_{-1}={1\over2}{c_{\a\b}}^\g\t_\g\partial_{\t_\a}\partial_{\t_\b},
\ea
\ba\label{1.4}
\D_{-3}={1\over3!}c_{\a\b\g}\partial_{\t_\a}\partial_{\t_\b}\partial_{\t_\g},
\ea
where
\ba
c_{\a\b\g}={c_{\a\b}}^\l g_{\l\g}\no
\ea
and
\ba
g_{\a\b}={c_{\a\l}}^\g{c_{\b\g}}^\l\no
\ea
is the Cartan-Killing metric tensor. The raising of the indices
\ba
\t^\a=g^{\a\b}\t_\b\no
\ea
is performed in the case of a semi-simple group with the help of the inverse
tensor $g^{\a\b}$ 
\ba
g^{\a\b}g_{\b\g}=\d^\a_\g.\no
\ea

The operator $\D_{+1}$ is proportional to the second term in a BRST charge
\ba
Q=\t^\a G_\a-{1\over2}\t^\a\t^\b{c_{\a\b}}^\g\partial_{\t^\g},\no
\ea
where $\t^\a$ and $\partial_{\t^\a}$ represent the operators for the ghosts 
and ghost momenta respectively and $G_\a$ are generators of the arbitrary 
Lie algebra. 
The operator $\D_{-1}$ related to the divergence of the vector field 
$\{\t_\a,A\}_1$
\ba\label{1.5}
\D_{-1}A=-{1\over2}\partial_{\t_\a}\{\t_\a,A\}_1
\ea
is proportional to the true $\D$-operator for the odd bracket (\ref{1.1}).
The operator $\D_{-1}$ (\ref{1.5}) determines the linear odd bracket 
(\ref{1.1}) as a deviation of the Leibniz rule under the usual multiplication
\begin{eqnarray}
\Delta_{-1}(A\cdot B)=(\Delta_{-1}A)\cdot B
+(-1)^{p(A)}A\cdot\Delta_{-1}B
+(-1)^{p(A)}\{A,B\}_1\ .\nonumber
\end{eqnarray}
and simultaneously satisfies the Leibniz rule with respect to the linear odd 
bracket composition
\ba
\D_{-1}(\{A,B\}_1)=\{\D_{-1}A,B\}_1+(-1)^{p(A)+1}\{A,\D_{-1}B\}_1,\no
\ea
where $p(A)$ is a Grassmann parity of the quantity $A$.

In the present paper we show that the operator $\D_{-3}$ (\ref{1.4}) is related
with the Grassmann-odd Nambu-like bracket on the Grassmann algebra. This 
bracket corresponds to an arbitrary Lie algebra.

\section{Nambu-like odd bracket}

By applying the operator $\D_{-3}$ (\ref{1.4}) to the usual product of two 
quantities $A$ and $B$, we obtain
\begin{eqnarray}\label{2.1}
\Delta_{-3}(A\cdot B)=(\Delta_{-3}A)\cdot B
+(-1)^{p(A)}A\cdot\Delta_{-3}B
+(-1)^{p(A)}\D(A,B),
\end{eqnarray}
where the quantity $\D(A,B)$ is
\ba\label{2.2}
\D(A,B)={1\over2}c_{\a\b\g}\left[\left(\partial_{\t_\a}\partial_{\t_\b}A
\right)\partial_{\t_\g}B+(-1)^{p(A)}\left(\partial_{\t_\a}A\right)
\partial_{\t_\b}\partial_{\t_\g}B\right].
\ea

By acting with the operator $\D_{-3}$ on the usual product of three quantities
$A$, $B$ and $C$, we come to the following relation:
\ba\label{2.3}
\D_{-3}(A\cdot B\cdot C)&=&\left(\D_{-3}A\right)\cdot B\cdot C+
(-1)^{p(A)}A\cdot \left(\D_{-3}B\right)\cdot C+
(-1)^{p(A)+p(B)}A\cdot B\cdot\D_{-3}C\cr&+&(-1)^{p(A)}\D(A,B)C+
(-1)^{p(A)p(B)+p(A)+p(B)}B\D(A,C)\cr&+&(-1)^{p(A)+p(B)}A\D(B,C)+
(-1)^{p(B)}\{A,B,C\}_1,
\ea
where the last term is the Grassmann-odd Nambu-like bracket
\ba\label{2.4}
\{A,B,C\}_1=c_{\a\b\g}\partial_{\t_\a}A\partial_{\t_\b}B\partial_{\t_\g}C
\ea
on the Grassmann algebra.

The divergence of the Nambu-like odd bracket (\ref{2.4}) is related with the 
quantity $\D(A,B)$ (\ref{2.2}) (see also \cite{sak})
\ba\label{2.5}
\D(A,B)={1\over2}\partial_{\t_\a}\{\t_\a,A,B\}_1,
\ea
whereas the multiplication on the Grassmann variable $\t^\a$ gives the linear
odd bracket (\ref{1.1})
\ba\label{2.6}
\{A,B\}_1=\t^\a\{\t_\a,A,B\}_1.
\ea
Note also the following relation between the operator $\D_{-3}$ (\ref{1.4})
and odd Nambu-like bracket (\ref{2.4}):
\ba\label{2.7}
\D_{-3}A={1\over3!}\partial_{\t_\a}\partial_{\t_\b}\{\t_\a,\t_\b,A\}_1.
\ea
For the operator $\D_{-3}$ (\ref{1.4}) there exists the following ``Leibniz
rule'' with respect to the odd Nambu-like bracket composition:
\ba\label{2.8}
\D_{-3}(\{A,B,C\}_1)=&-&\{\D_{-3}A,B,C\}_1+(-1)^{p(A)}\{A,\D_{-3}B,C\}_1\cr
&-&(-1)^{p(A)+p(B)}\{A,B,\D_{-3}C\}_1\cr&+&c_{\a\b\g}\Bigl[(-1)^{p(A)+1}
\D\left(\partial_{\t_\a}A,\partial_{\t_\b}B\right)\partial_{\t_\g}C\cr
&+&(-1)^{p(A)+p(B)}\partial_{\t_\a}A\D\left(\partial_{\t_\b}B,
\partial_{\t_\g}C\right)\cr&+&(-1)^{p(A)p(B)+1}\partial_{\t_\b}B
\D\left(\partial_{\t_\a}A,\partial_{\t_\g}C\right)\Bigr].
\ea

It follows from the expression (\ref{2.4}) for the odd Nambu-like bracket 
the Grassmann parity
\ba\label{2.9}
p(\{A,B,C\}_1)=p(A)+p(B)+p(C)+1\pmod2,
\ea
symmetry properties
\ba\label{2.10}
\{A,B,C\}_1=-(-1)^{[p(A)+1][p(B)+1]}\{B,A,C\}_1
=-(-1)^{[p(B)+1][p(C)+1]}\{A,C,B\}_1
\ea
and Jacobi type identity 
\ba\label{2.11}
\{\{A,B,C\}_1,D,E\}_1&+&(-1)^{[p(A)+1][p(B)+p(C)+p(D)+p(E)]}
\{\{B,C,D\}_1,E,A\}_1\cr&+&(-1)^{[p(A)+p(B)][p(C)+p(D)+p(E)+1]}
\{\{C,D,E\}_1,A,B\}_1\cr&+&(-1)^{[p(D)+p(E)][p(A)+p(B)+p(C)+1]}
\{\{D,E,A\}_1,B,C\}_1\cr&+&(-1)^{[p(E)+1][p(A)+p(B)+p(C)+p(D)]}
\{\{E,A,B\}_1,C,D\}_1\cr&+&(-1)^{[p(D)+p(E)][p(A)+p(B)+p(C)+1]+p(B)[p(A)+1]
+p(A)}\cr&\times&\{\{D,E,B\}_1,A,C\}_1\cr&+&(-1)^{[p(D)+1][p(A)+p(B)+p(C)]
+p(D)}
\{\{D,A,B\}_1,C,E\}_1\cr&+&(-1)^{[p(A)+1][p(B)+p(C)+p(D)+p(E)]+[p(D)+1]p(E)
+p(D)}\cr&\times&\{\{B,C,E\}_1,D,A\}_1\cr&+&(-1)^{[p(B)+1][p(C)+p(D)+p(E)]
+p(B)}
\{\{A,C,D\}_1,E,B\}_1\cr&+&(-1)^{[p(B)+1][p(C)+p(D)+p(E)+1]+[p(D)+1][p(E)+1]}
\cr&\times&\{\{A,C,E\}_1,D,B\}_1=0.
\ea
Note that the structure of (\ref{2.11}) is different from the one for the
fundamental identity \cite{t}.

\section{Conclusion}

Thus, we constructed the Grassmann-odd Nambu-like bracket which corresponds to 
the arbitrary Lie algebra and are realized on the Grassmann
algebra. The main properties of this bracket are also given.

\acknowledgments

One of the authors (V.A.S.) thanks the administration of the Office of 
Associate and Federation Schemes of the Abdus Salam ICTP for the kind 
hospitality at Trieste where this work has been started.

\end{document}